\documentclass[11pt]{article}

\pdfoutput=1

\usepackage{times}

\usepackage[authoryear,round,sort&compress]{natbib}

\usepackage{amsmath}
\usepackage{amsfonts}
\usepackage{amsthm}

\usepackage{graphicx}

\usepackage[bookmarks=false,hypertex,colorlinks=false,citebordercolor={1 1 1},filebordercolor={1 1 1},linkbordercolor={1 1 1},menubordercolor={1 1 1},urlbordercolor={1 1 1},runbordercolor={1 1 1}]{hyperref}
\usepackage{bookmark}

\theoremstyle{plain}

\newtheorem{proposition}{Proposition}
\theoremstyle{definition}

\newtheorem{assumption}{Assumptions}

\theoremstyle{remark}

\newcommand\Prob{\text{P}}  

\begin{document}

\title{Predicting Sequences of Progressive Events Times with Time-dependent
Covariates}

\author{Song Cai, James V. Zidek\thanks{Department of Statistics, the
University of British Columbia, 333-6356 Agricultural Road, Vancouver, BC, V6T
1Z2}, Nathaniel Newlands\thanks{Environmental Health, Agriculture and
Agri-Food Canada, 5403 - 1st Avenue S., P.O. Box 3000, Lethbridge, Alberta,
Canada}}

\maketitle

\begin{abstract}

This paper presents an approach to modeling progressive event-history data
when the overall objective is prediction based on time-dependent covariates.
This approach does not model the hazard function directly. Instead, it models
the process of the state indicators of the event history so that the
time-dependent covariates can be incorporated and predictors of the future
events easily formulated. Our model can be applied to a range of real-world
problems in medical and agricultural science.

\end{abstract}

\section{Introduction}

This paper presents a new theory for event history processes that involve a
sequence of irreversible, progressive events and associated external
time-dependent covariates, i.e. covariates not influenced by the occurrence of
the events of central interest (\citealt{Kalbfleisch2002}). These covariates are
known up to times on a discrete scale, say daily scale, for example. Such
events signal changing conditions, which may point to the need for strategic
actions that reduce risk associated with those processes, for example cancer
progression or survival of wine-grape perennial crops. By modeling how such
event sequences change in relation to time-dependent covariates, useful
information may be provided to those involved in assessing best therapeutic
intervention responses, or environmental impacts.

General event-history data have been well studied.  As examples:
\citet{Weiss1965} and \citet{Lagakos1978} considered semi-Markov models;
\citet{Hougaard2000} described a broad range of Markov models;
\citet{Cook2007} presented two broad approaches for recurrent event data:
modeling the counts of events in a time interval and modeling the gap time
between two events; \cite{Aalen2008} described approaches based on counting
processes. Many of these approaches can be used to analyze progressive events
data. However, when a time-dependent covariate is present, the problem becomes
thorny, especially if the main objective of the analysis is prediction.

When a single event is under consideration and a time-dependent covariate is
present, the usual practice is to apply the Cox model \citep{Cox1972} or a
parametric proportional hazards model \citep{Collett2003}. The advantage of
the Cox model is that if the hazard function is only related to the covariate
evaluated at the current time, then we can plug that covariate value into an
expression for a partial likelihood function, regardless of the values of the
covariate at other time points. However, this causes a loss of efficiency,
since the information contained in the covariate between the gap times of
events are not used. \citet{Cox1972} argued that the loss of efficiency is not
much unless either:(1) the model parameter is far from zero; (2) censoring is
strongly dependent on covariates; or (3) there are strong time trends in the
covariates. While the first two issues may not concern us, the third is
crucial for phenological data since the associated climate variables usually
have strong seasonality (and will exhibit as a dominant local trend between
event-times within a season). On the other hand, the Cox model is not suitable
for prediction, since it does not extrapolate beyond the last observation. A
parametric proportional hazards model might be a good choice for prediction.
But it requires explicit distributional assumptions for the time-to-event,
which may be mis-specified. Also, if the hazard ratio is related to the
covariate evaluated at several time points at and prior to the current time,
the likelihood function may involve a complicated integration.

When multiple events are of interest, to deal with time-dependent covariates,
the usual approach is to apply a Cox model for each event where time-dependent
covariates are present (e.g. \citealp{Hougaard1999}), or use a parametric
model to model the hazard rate and to incorporate the covariates just as in
the parametric proportional hazards model (e.g. \citealp{Cook2007}). These
approaches induce similar problems to those in the single event case.

In this paper, we introduce an approach based on modeling the process of event
state indicator. In this approach, all the available information contained in
time-dependent covariates can be easily incorporated in the likelihood
function, and the construction of a predictor is straightforward. Also, this
approach does not impose strong distributional assumption on times to events.

The paper is organized as follows. Section \ref{sec:single} presents a model
for a single event. There our basic assumptions are introduced and estimation
and prediction procedures for the model are described. Also, the estimation
for the case of non-informative right censored response is considered. Section
\ref{sec:multiple} presents a model for sequential events, which is an
extension of our model for single event but with a few additional assumptions.
In section \ref{sec:example}, we test our model for single event by applying
it to the blooming event of pear trees. There a cross validation procedure is
used to evaluate our prediction of future events. The uncertainty associated
with the prediction is also assessed. The final section summarizes our
methods, and gives pointers to possible future work.

\section{Model for a single event} \label{sec:single}

This section concerns the case of a single phenological outcome called an
``event'', for example  ``death''.  The data consist of the times to the
occurrence of that outcome for $N$ experimental subjects,  $i=1,\;\cdots,\;N$.

\subsection{Basic setup} \label{subsec:datasingle}

In the sequel, upper case letters denote random variables and 
lower case ones, their realized values.

We adopt the following assumption in this section: 
\begin{assumption} \label{ass:1single}
Only one event can occur for each individual, and once it has occurred, it
remains in the ``occurred'' state thereafter.
\end{assumption}

We assume a discrete time scale with a well-defined origin $t_0$, that we take
to be $t_0=0$ without loss of generality. For individual $i$, let $T_i$ denote
the random time to occurrence of the event. At each time point $t=0,
\:1,\:\cdots$, classify the state of the event for each individual as
``occurred'' or ``not occurred''. At time $t$, let $Y_{i,\,t}$ denote this
state, being 1 or 0 according as the event has ``occurred'' or not.  Then,
time to event $T_i$ and state indicator $Y_{i,\,t}$ have the following
relationship:
\begin{equation}
  Y_{i,\,0}=0, \; Y_{i,\,1}=0, \; \cdots, \; Y_{i,\,(T_i-1)}=0, \;
  Y_{i,\,T_i}=1, \; Y_{i,\,\left(T_i+1\right)}=1, \; \cdots \;,
  \label{relation2}
\end{equation}
where $Y_{i,\,t}$ is 1 for all $t \ge T_i$ by Assumptions \ref{ass:1single}.

Associated with each individual is a time-dependent covariate vector, which is
observed on the same discrete time scale. Denote its value at time $t$
($t=\cdots, \;-1,\;0, \:1,\:\cdots$) by $X_{i,\,t}$. Note that a fixed
covariate is a special time-dependent one and so is subsumed by our theory.
For individual $i$, we further denote the covariate process evaluated at all
time points, i.e. $\{\cdots\,\;X_{i,\,-1}=x_{i,\,-1}, \;X_{i,\,0}=x_{i,\,0},
\;X_{i,\,1}=x_{i,\,1}, \;\cdots\}$, as $X_{i,\,t^\prime\in \mathbb{Z}}$.
Similarly, we write $Y_{i,\,0:t}$ as the set of state indicators $Y_{i,\,t}$
evaluated from time origin 0 to time point $t$ ($t=0, \:1,\:\cdots$), i.e. $\{
Y_{i, \, 0}=y_{i, \,0}, \;\cdots,\;Y_{i,\,t}=y_{i,\,t}\}$.

\subsection{Probability model} \label{subsec:probsingle}

The conditional probability distribution of $Y_{i,\,0:t}$ given $X_{i,\,t^\prime\in
\mathbb{Z}}$ is
\begin{equation}
  \Prob \left( Y_{i,\,0:t} \left| X_{i,\,t^\prime\in \mathbb{Z}}
    \right.\right) = \Prob\left(Y_{i,\,0}=y_{i,\,0} \left| X_{i,\,t^\prime\in
    \mathbb{Z}} \right. \right) \prod_{s=1}^{t} \Prob\left(Y_{i,\,s}=y_{i,\,s}
    \left| Y_{i,\,0:\left(s-1\right)}, \;X_{i,\,t^\prime\in \mathbb{Z}}
    \right. \right) \;,
    \label{eq:singlefullp}
\end{equation}
where $\Prob\left(\cdot\right)$ is the probability set function. This
expression can be simplified using the following result:
\begin{proposition} \label{propMC}
  For each individual $i$ and single event in Assumption
  \ref{ass:1single}, conditional on $X_{i,\,t^\prime\in \mathbb{Z}}$, the
  stochastic process $\left\{Y_{i,\,t}: t=0,\;1, \;\cdots \right\}$ is a first
  order Markov chain\index{Markov chain}, i.e.
  \begin{equation}
    \Prob \left(Y_{i,\,t}=y_{i,\,t} \left| Y_{i,\,0:\left(t-1\right)}, \;
    X_{i,\,t^\prime\in \mathbb{Z}}\right.\right) = \Prob
    \left(Y_{i,\,t}=y_{i,\,t} \left|
    Y_{i,\,\left(t-1\right)}=y_{i,\,\left(t-1\right)}, \; X_{i,\,t^\prime\in
    \mathbb{Z}}\right.\right) \ ,
  \end{equation}
  for all $t=1,\;2, \;\cdots$ and $y_{i,\,t}\in \left\{0,\;1\right\}$.
\end{proposition}
\begin{proof}
  In the proof everything will be conditional on $X_{i,\,t^\prime\in
  \mathbb{Z}}$ and is omitted for simplicity. Since for each individual $i$
  and all $t=0,\;1,\;\cdots$, $Y_{i,\,t}$ is a binary random variable, it
  suffices to separately consider only two cases, $Y_{i,\,\left(t-1\right)}=0$
  and $Y_{i,\,\left(t-1\right)}=1$. Firstly, $Y_{i,\,\left(t-1\right)}=0$
  implies that $Y_{i,\,0}=0, \;\cdots$, and $Y_{i,\,\left(t-2\right)}=0$,
  making $\{Y_{i,\,0}=0, \;\cdots, Y_{i,\,\left(t-2\right)}=0,
  Y_{i,\,\left(t-1\right)}=0\}$ the only possible probability event for
  $Y_{i,\,0:\left(t-1\right)}$ and thus equivalent to
  $\{Y_{i,\,\left(t-1\right)}=0\}$. 
  
  Secondly, for the type of single event under consideration, if for some
  $t^\prime>0$, $Y_{i,\,\left(t^\prime-1\right)}=1$, then
  $Y_{i,\,\left(t-1\right)}=1$ for all $t\ge t^\prime$. Thus, when
  $Y_{i,\,\left(t-1\right)}=1$ ($t>0$), we have
  \begin{align}
    \Prob \left(Y_{i,\,t}=y_{i,\,t} \left|
    Y_{i,\,0:\left(t-1\right)}\right.\right)
    =\Prob \left(Y_{i,\,t}=y_{i,\,t} \left|
    Y_{i,\,0:\left(t-2\right)}, \:Y_{i,\,\left(t-1\right)}=1\right.\right)
    =1,
  \end{align}
\end{proof}
which completes the proof.$\bullet$

Equation (\ref{eq:singlefullp}) then simplifies to
\begin{equation}
  \Prob \left( Y_{i,\,0:t} \left| X_{i,\,t^\prime\in \mathbb{Z}} \right.\right) =
    \Prob\left(Y_{i,\,0}=y_{i,\,0} \left| X_{i,\,t^\prime\in \mathbb{Z}} \right.
    \right) \prod_{s=1}^{t} \Prob\left(Y_{i,\,s}=y_{i,\,s} \left|
    Y_{i,\,\left(s-1\right)}=y_{i,\,\left(s-1\right)}, \;X_{i,\,t^\prime\in
    \mathbb{Z}} \right. \right) \;.
    \label{eq:singlefullps}
\end{equation}
For individual $i$,the previous equation and (\ref{relation2}) imply that the
conditional probability that the event occurs at time $t_i$, given all the
covariate values $X_{i,\,t^\prime\in \mathbb{Z}}$ is 
\begin{align}
  \Prob \left( T_i=t_i \left| X_{i,\,t^\prime\in \mathbb{Z}} \right. \right) 
    &= \Prob  \left( Y_{i,\,0}=0, \; Y_{i,\,1}=0, \; \cdots, \;
    Y_{i,\,(t_i-1)}=0, \; Y_{i,\,t_i}=1 \left| X_{i,\,t^\prime\in \mathbb{Z}}
    \right. \right) \notag \\
    &= \Prob\left(Y_{i,\,0}=0 \left| X_{i,\,t^\prime\in \mathbb{Z}} \right.
    \right) \cdot \left[ \prod_{s=1}^{t_i-1} \Prob\left(Y_{i,\,s}=0 \left|
    Y_{i,\,\left(s-1\right)}=0, \;X_{i,\,t^\prime\in \mathbb{Z}} \right.
    \right) \right] \cdot \notag \\
    &\qquad \Prob \left( Y_{i,\,t_i}=1 \left|
    Y_{i,\,\left(t_i-1\right)}=0, \;X_{i,\,t^\prime\in \mathbb{Z}}
    \right.\right) \;.
    \label{eq:likeraw}
\end{align}
Now we are ready to build a regression model based on this probability model.

\subsection{Regression model} \label{subsec:regModelSingle}

Assume that the occurrences of the events of different individuals are
independent realizations from the same population. We require additional
assumptions about the relationship of the occurrence of the event and
covariate to limit the total number of parameters. In Equation
(\ref{eq:likeraw}), the probability of an event occurring at time $t_i$ is
conditioned on covariate values evaluated at all discrete time points $\cdots,
\;-1,\;0, \;1, \;\cdots$. In real applications, the occurrence of an event
usually only depends on the covariate values at and prior to the occurrence
time. Furthermore, in some situations, we may assume that at a time point $t
\ge 0$, the state of the event mainly depends on the covariate values at the
current and several previous times, or some weighted average of them. In
practice, we want to make some reasonable assumptions so that the total number
of covariates (and consequently the total number of parameters in the
regression model) is limited, and the number of covariates does not change
over time.

For the purpose of illustration, simply assume that for individual $i$ at time
point $t$, the state indicator $Y_{i,\,t}$ is only related to the covariate
values evaluated from time $t-K$ to $t$, i.e. $\left\{
X_{i,\,\left(t-K\right)}, \;\cdots,\; X_{i,\,t}\right\}$, where $K$ is
constant. Now, for individual $i$ at time $t$, no matter if $X_{i,\,t}$
is a vector or not, the total number of covariate values that are related to
$Y_{i,\,t}$ is finite and fixed, and we will put them together as a vector
denoted by $\mathcal{X}_{i,\,t}$. Then in Equation (\ref{eq:likeraw}), term
$X_{i,\,t^\prime\in \mathbb{Z}}$ on the right hand side (RHS) can be replaced
by $\mathcal{X}_{i,\,t}$.

We further assume that time origin $0$ is the earliest time an event can
occur, otherwise the data is not useful for studying the probability of the
occurrence of the event. Then we have
\begin{equation}
  \Prob\left(Y_{i,\,0}=y_{i,\,0} \left| \mathcal{X}_{i,\,0} \right. \right) =
  \Prob\left(Y_{i,\,0}=y_{i,\,0} \left| Y_{i,\,-1}=0,\  \mathcal{X}_{i,\,0}
  \right. \right) \ .
\end{equation}

By virtue of the Markov property of $\left\{Y_{i,\,t}: t=0,\;1, \;\cdots
\right\}$, for modeling $\Prob \left( T_i=t_i \left| X_{i,\,t^\prime\in
\mathbb{Z}} \right. \right)$, it suffices to model
$\Prob\left(Y_{i,\,t}=y_{i,\,t}\left| Y_{i,\,\left(t-1\right)}=0,
\;\mathcal{X}_{i,\,t} \right. \right)$ for $t=0,\;\cdots,\;t_i$ and
$y_{i,\,t}\in\{0,\;1\}$. Write
\begin{equation}
  \Prob_{i,\,t}\equiv\Prob\left(Y_{i,\,t}=1\left| Y_{i,\,\left(t-1\right)}=0,
    \;\mathcal{X}_{i,\,t} \right. \right) \ .
\end{equation}
Then since $y_{i,\,t}\in\{0, \, 1\}$, we have
\begin{equation}
  \Prob\left(Y_{i,\,t}=y_{i,\,t} \left| Y_{i,\,\left(t-1\right)}=0,
  \;\mathcal{X}_{i,\,t} \right. \right) = \Prob_{i,\,t}^{y_{i,\,t}}
  \left(1-\Prob_{i,\,t} \right)^{1-y_{i,\,t}} \ .
\end{equation}
For each fixed individual $i$, $\Prob_{i,\,t}$ is a function of $t$ and
$\mathcal{X}_{i,\,t}$. Now, to build a regression model, we will choose a
useful explicit form for this function with unknown parameters, and carry out
statistical inference for these parameters.

If we want to restrict the functional form of $\Prob_{i,\,t}$ to a linear
function of the parameters, then for individual $i$, at each time point
$t=0,\;\cdots,\;t_i$, we may consider a linear regression model for binary
events. Consider a monotonic link function $g:\left(0,\,1\right)
\rightarrow\left(-\infty, \infty \right)$ (e.g., $g$ could be the logit or
probit function).  We assume that $g\left( \Prob_{i,\,t} \right)$ equals a
linear function of the covariate vector $\mathcal{X}_{i,\,t}$, i.e.
\begin{equation}
  g\left( \Prob_{i,\,t} \right) = \beta_t^T\mathcal{X}_{i,\,t} \ ,
  \label{eq:simplereg}
\end{equation}
where $\beta_t$ is a parameter vector which remains the same across different
individuals $i$, but may vary with time $t$. The superscript $T$ stands for the
transpose of a vector or a matrix.

By Equation (\ref{eq:likeraw}) -- (\ref{eq:simplereg}), we have
\begin{align}
  \Prob \left( T_i=t_i \left| X_{i,\,t^\prime\in \mathbb{Z}} \right. \right) 
    &= g^{-1}\left(\beta_t^T\mathcal{X}_{i,\,t_i}\right)\prod_{s=0}^{t_i-1}
    \left( 1-g^{-1}\left(\beta_t^T\mathcal{X}_{i,\,s}\right)\right) \ ,
    \label{eq:regsingle}
\end{align}
where $g^{-1}$ is the inverse function of $g$. To achieve computational
tractability, we take $\beta_t$ to be a constant vector over time, so the
subscript $t$ of $\beta_t$ in the above equation can be omitted. Under the
independence assumption, the likelihood function of the data is
\begin{equation}
  L \left(\beta \right) = \prod_{i=1}^{N} \left[
    g^{-1}\left(\beta^T\mathcal{X}_{i,\,t_i}\right)\prod_{s=0}^{t_i-1}
    \left( 1-g^{-1}\left(\beta^T\mathcal{X}_{i,\,s}\right)\right)
    \right] \ ,
    \label{eq:liksingle}
\end{equation}
One can now proceed with maximum likelihood (ML) or Bayesian
methods to estimate parameters.

\subsection{Non-informative right censoring}\index{censoring!non-informative
right}
\label{sec:censoring}

If the event has not occurred for an individual by the end of the study or an
individual left the study before the event occurs, we get a right-censored
observation. In this paper, we consider non-informative right censoring, i.e.
the time to the event is independent of the censoring mechanism. The
methodology is derived from \citet{Collett2003}.

Here, when writing the conditional probability of an event occurring at some
time point given covariate values $X_{i,\,t^\prime\in \mathbb{Z}}$, we will
omit the conditioning variable. All the probability expressions in this
section are then conditioned on $X_{i,\,t^\prime\in \mathbb{Z}}$.

For each individual $i=1,\;\cdots,\;N$, we have an observed time $t_i$,
which is either an event-time, or a right censoring time. We denote this
observation as a random variable $\tau_i$. Then the value of $\tau_i$ is
$t_i$. Now, for individual $i$, let $\delta_i$ be an indicator which
takes values 1 or 0, according as we observe the event or not because it
is right censored. By the non-informative censoring assumption, we can assume that
each individual $i$ is associated with two independent random variables: event
time $T_i$ and censoring time $C_i$. If the observation for individual $i$ is
censored, we have
\begin{equation}
  C_i < T_i \ \text{and } \tau_i=C_i, \text{ when } \delta_i=0 \ ,
\end{equation}
otherwise, we have
\begin{equation}
  C_i > T_i \ \text{and } \tau_i=T_i, \text{ when } \delta_i=1 \ .
\end{equation}

Now, it is easy to see that $\tau_i=\min\left(T_i, \;C_i\right)$, and
\begin{align}
  \Prob \left( \tau_i=t, \: \delta_i=0\right) 
    = \Prob \left(C_i=t,\:T_i>t \right)
    = \Prob \left(C_i=t\right) \Prob \left( T_i>t \right) \ ,
\end{align}
where the second equality holds because of the non-informative censoring
assumption. Similarly, we have
\begin{align}
  \Prob \left( \tau_i=t, \: \delta_i=1\right) 
    = \Prob \left(T_i=t,\:C_i>t \right)
    = \Prob \left(T_i=t\right) \Prob \left( C_i>t \right) \ .
\end{align}
The likelihood function for the observations $t_1,\:\cdots,\:t_N$ then is
\begin{align}
  L &= \prod_{i=1}^N \Prob \left( \tau_i=t_i, \: \delta_i\right) \notag \\
    &=\prod_{i=1}^N \left( \Prob \left(C_i=t_i\right) \Prob \left( T_i>t_i \right)
    \right)^{1-\delta_i} \left( \Prob \left(T_i=t_i\right) \Prob
    \left( C_i>t_i \right) \right)^{\delta_i} \notag \\
    &=\left[\prod_{i=1}^N \Prob \left(C_i=t_i\right)^{1-\delta_i}\Prob
    \left( C_i>t_i \right)^{\delta_i}\right] \left[\prod_{i=1}^N \Prob
    \left(T_i=t_i\right)^{\delta_i}\Prob \left( T_i>t_i
    \right)^{1-\delta_i}\right] \ .
\end{align}
By the non-informative censoring assumption, term $\left[\prod_{i=1}^N
\Prob \left(C_i=t_i\right)^{1-\delta_i}\Prob \left( C_i>t_i
\right)^{\delta_i}\right]$ does not involve parameters that are related to the
distribution of event-time $T_i$. Therefore, to find the maximum likelihood
estimator (MLE) of the model parameters, it suffices to maximize the following
function
\begin{equation}
  L^\prime \left( \beta \right) = \prod_{i=1}^N \Prob
    \left(T_i=t_i\right)^{\delta_i}\Prob \left( T_i>t_i
    \right)^{1-\delta_i} \ ,
    \label{eq:censorlik}
\end{equation}
and
\begin{equation}
  \hat{\beta}_{MLE} = \text{Argmax} L^\prime \left( \beta \right) \ .
  \label{eq:censormle}
\end{equation}

Term $\Prob \left(T_i=t_i\right)$ in Equation (\ref{eq:censorlik}) is
given by Equation (\ref{eq:regsingle}) (note that the conditioning variable
$X_{i,\,t^\prime\in \mathbb{Z}}$ has been omitted in the current expressions),
while term $\Prob \left( T_i>t_i \right)$ can be calculated as follows:
\begin{align}
  \Prob \left( T_i>t_i \right) &= \Prob \left( Y_{i,\,0}=0,
    \;Y_{i,\,1}=0,\;\cdots, \;Y_{i,\,t_i}=0\right)
    = \prod_{s=0}^{t_i} \left(
    1-g^{-1}\left(\beta^T\mathcal{X}_{i,\,s}\right)\right) \ .
\end{align}
Then we can re-write Equation (\ref{eq:censorlik}) as
\begin{align}
  L^\prime \left( \beta \right) = \prod_{i=1}^N
    \left[
    g^{-1}\left(\beta^T\mathcal{X}_{i,\,t_i}\right)\prod_{s=0}^{t_i-1}
    \left( 1-g^{-1}\left(\beta^T\mathcal{X}_{i,\,s}\right)\right)
    \right]^{\delta_i}
    \left[\prod_{s=0}^{t_i} \left(
    1-g^{-1}\left(\beta^T\mathcal{X}_{i,\,s}\right)\right)\right]^{1-\delta_i}
    \ .
    \label{eq:censorlikmodel}
\end{align}
Now, we can easily estimate $\beta$ using Equation (\ref{eq:censormle}) if
it is assumed constant over time.

\subsection{Prediction} \label{sec:2prediction}

To use the regression model to predict the time to a future event, we must
know the future values of time-dependent covariates in advance. However,
generally we will not know them and hence must predict them. We therefore
assume that their predictive distribution is available in order to make 
progress on this problem.

With that understanding and time origin 0, suppose the current time is
$t_c\ge0$. For a new individual, one whose data were not used for parameter
estimation and whose event-time is unknown, suppose the event has not
occurred up to $t_c$. This subsection presents a predictor for the event-time
of this individual, denoted by $T^*$, with corresponding state indicator
$Y^*_t$ at time $t\ge 0$. 

To construct that predictor, we denote the covariate vector for this
individual evaluated at time $t$ as $X^*_t$. Similarly we write the ``new
individual version'' of $\mathcal{X}_{i,\,t}$ as $\mathcal{X}^*_{t}$. Since we
know the covariate values up to time $t_c$, $\mathcal{X}^*_{t}$ may be
decomposed into two vectors: one vector $\mathcal{X}^*_{t,\,obs}$ consists of
covariates values evaluated from time 0 to time $t_c$, which we observed
exactly, and the other, $\mathcal{X}^*_{t,\,pred}$, consists of predicted
covariates values from $t_c+1$ to $t$, whose predictive distributions are
given by another model. Furthermore, denote the estimated parameter vector as
$\hat{\beta}$, and the covariates and state indicator used to estimate
$\hat{\beta}$ as $X^{train}$ and $Y^{train}$ respectively.

If we knew the true value of $\beta$, Equation (\ref{eq:regsingle}) would
imply the predictive distribution of bloom time $T^*$.  In other words the
probability of the event occurring at time $t_c+K$ for any $K\ge1$ given
$\mathcal{X}^*_{t,obs}$, the observed covariate values for the new individual,
would be
\begin{align}
  &\Prob_{\beta}\left( T^*=t_c+K \left| \mathcal{X}^*_{t_c+K,\,obs} \right.
    \right)
    \notag \\
    =&
    \int \Prob_{\beta}\left( T^*=t_c+K \left| \mathcal{X}^*_{t_c+K,\,obs}, \;
    \mathcal{X}^*_{t_c+K,\,pred} \right. \right) d
    \Prob \left(\mathcal{X}^*_{t_c+K,\,pred} \right)
    \notag \\
    =&
    \int g^{-1}\left(\beta^T\mathcal{X}^*_{t_c+K}\right)
    \prod_{s=1}^{K-1}
    \left( 1-g^{-1}\left(\beta^T\mathcal{X}^*_{t_c+s}\right)\right)
    d \Prob \left(\mathcal{X}^*_{t_c+K,\,pred} \right)
    \ .
    \label{eq:preddist}
\end{align}
We attach a subscript $\beta$ on that probability function to emphasize we are
using the true parameter values. The problem is that we do not know the true
$\beta$. So we replace $\beta$ by $\hat{\beta}$ in Equation
(\ref{eq:preddist}) to estimate the predictive distribution of the event-time
$T^*$ as:
\begin{align}
  &\Prob_{\hat{\beta}}\left( T^*=t_c+K \left |
    \mathcal{X}^*_{t_c+K,\,obs} \right. \right)
    \notag \\
    =&
    \int g^{-1}\left(\hat{\beta}^T\mathcal{X}^*_{t_c+K}\right)
    \prod_{s=1}^{K-1}
    \left( 1-g^{-1}\left(\hat{\beta}^T\mathcal{X}^*_{t_c+s}\right)\right)
    d \Prob \left(\mathcal{X}^*_{t_c+K,\,pred} \right)
    \ .
    \label{eq:preddistplugin}
\end{align}
If the predictive distribution of $\mathcal{X}^*_{t_c+K,\,pred}$ is given by
another model, the integral in this equation may be calculated by the Monte Carlo
(MC) algorithm. Generate a sample of large size $L$ from the
distribution of $\mathcal{X}^*_{t_c+K,\,pred}$, and denote the sample points
as $\mathcal{X}^*_{t_c+K,\,pred}\left(l \right)$ ($l=1,\,\cdots,\,L$). Then we
may approximate the predictive probabilities by
\begin{equation}
  \Prob_{\hat{\beta}} \left( T^*=t_c+K \lvert \mathcal{X}^*_{t, obs}\right)
  \approx \frac{1}{L}\sum_{l=1}^L \Prob_{\hat{\beta}}\left( T^*=t_c+K \left|
    \mathcal{X}^*_{t_c+K,\,obs}, \; \mathcal{X}^*_{t_c+K,\,pred}\left(l
    \right) \right. \right) \ .
    \label{eq:predictivedMC}
\end{equation}

This ``plug-in'' approach for predictive distribution is generally criticized
as failing to take into account the uncertainty of the unknown parameter. But,
if one takes the Bayesian approach, the uncertainty of the unknown parameter
is incorporated in a natural way. Suppose in an estimation procedure, one
takes the Bayesian approach and gets $\Prob\left(\beta \lvert X^{train}, \;
Y^{train}\right)$, the posterior distribution of $\beta$. Then the predictive
distribution of $T^*$ is:
\begin{align}
  &\Prob \left( T^*=t_c+K \left| \mathcal{X}^*_{t_c+K, obs}, \;
    X^{train}, \; Y^{train} \right. \right)
    \notag \\
    =&
    \int \int \Prob\left( T^*=t_c+K \left| \mathcal{X}^*_{t_c+K, obs}, \;
    \mathcal{X}^*_{t_c+K, pred}, \; \beta, \; X^{train}, \; Y^{train} \right.
    \right)
    \notag \\
    & \qquad \qquad
    d \Prob \left(\beta \lvert X^{train}, \; Y^{train} \right)
    d \Prob \left(\mathcal{X}^*_{t_c+K, pred} \right)
    \ .
    \label{eq:preddistbayesian}
\end{align}
One may expect that the Bayesian approach will in general be superior to the
``plug-in'' approach in terms of prediction. However, \citet{Smith1998} showed
that for many models, when assessed from the point of view of mean squared
error of predictive probabilities, the ``plug-in'' approach is better than the
Bayesian approach in the extreme tail of the distribution. It is not directly
clear if this argument fits our model, but the point here is that we think
both approaches make sense.

\section{Model for multiple events} \label{sec:multiple}

\subsection{Basic setup} \label{sec:notationmultp}

Suppose there are $N$ individuals, and $S \ge1$ different events may
occur to each individual. We make the following assumption:
\begin{assumption} \label{ass:2mult}
For each individual, the $S$ events have the following properties:
\begin{enumerate}

  \item They occur in a fixed time order;

  \item For an event to occur, all the events prior to it must have occurred.

  \item For a fixed individual, no two different events occur at the same time point.

\end{enumerate}
\end{assumption}

By these assumptions, we can label each event by the time order in which it
occurs, using the symbol $s=1,\:,\cdots,\:S$. When we talk about the
occurrence of the $s^{th}$ event, all the previous events from the $1^{st}$ to
the $\left(s-1\right)^{st}$ must have occurred.

Now, for an individual $i$, there are $S+1$ states: no events have occurred,
the first event has occurred but the second hasn't and so on to the last event
has occurred, i.e. all $S$ events have occurred. We will denote these states
by $0, \;1, \;\cdots, \;S$, respectively. For the $i^{th}$ individual, we will
denote the random variable for the time to the $s^{th}$ event as $T_{i,\,s}$,
and denote its value as $t_{i,\,s}$,  $s=1,\;\cdots,\;S$. We also create a
state indicator $Y_{i,\,t}$ with $Y_{i,\,t}=l \in \{0, \;1, \;\cdots, \;S\}$
indicating that the individual $i$ is in the $l^{th}$ state. For the $i^{th}$
individual, starting from the time origin 0, we consider discrete time points
$0, \,1, \,\cdots, t_{i,\,1}, \cdots, t_{i,\,2}, \cdots, t_{i,\,S}$. The time
origin 0 satisfies $0 \le t_{i,\,1}$. The value of $Y_{i,\,t}$ can only be $l$
or $l+1$ when $Y_{i,\,t-1}=l\in \{0, \;1, \;\cdots, \;S-1\}$. Also,
$Y_{i,\,t}=S$ for all $t\ge t_{i,\,S}$. Then, the event-times $\{T_{i,\,s}\}$
and state indicators $\{Y_{i,\,t}\}$ have the following relationship:
\begin{eqnarray}\nonumber
  Y_{i,\,0}=0, \:\cdots, \: Y_{i,\,t_{i,\,1}}=1, \: Y_{i,(\,t_{i,\,1}+1)}=1, 
  \:\cdots, \: Y_{i,\,( t_{i,\,S}-1) }=S-1, \\
  Y_{i,\,t_{i,\,S}}=S, \: Y_{i,\,(t_{i,\,S}+1)}=S, \: \cdots
\end{eqnarray}
Furthermore, assume that at each discrete time point, we observe a covariate
vector $X_{i,\,t}$.

With the above notation, we will continue to let $Y_{i,\,0:t}$ denote $\{
Y_{i, \, 0}=y_{i, \,0}, \;\cdots,\;Y_{i,\,t}=y_{i,\,t}\}$, and
$X_{i,\,t^\prime\in \mathbb{Z}}$ denote $\{\cdots\,\;X_{i,\,-1}=x_{i,\,-1},
\;X_{i,\,0}=x_{i,\,0}, \;X_{i,\,1}=x_{i,\,1}, \;\cdots\}$, as we did above for a 
single event.

\subsection{Probability model}

For multiple progressive events that satisfy Assumptions \ref{ass:2mult} and 
each individual $i$, the conditional probability of $Y_{i,\,0:t}$ given
$X_{i,\,t^\prime\in \mathbb{Z}}$ still satisfies Equation
(\ref{eq:singlefullp}). However, the stochastic process $\left\{Y_{i,\,t}:
t=0,\;1, \;\cdots \right\}$ is no longer necessarily a first-order Markov
chain\index{Markov chain}. Instead, the following result holds with $ \; l=1,\;\cdots,\;S-1$:
\begin{align}
  &\Prob \left(Y_{i,\,t}= y_{i,\,t} \left| Y_{i,\,0:\left(t-1\right)}, \;
    X_{i,\,t^\prime\in \mathbb{Z}}\right.\right) \notag \\ = &\left\{
    \begin{array}{ll} \Prob \left(Y_{i,\,t}= y_{i,\,t} \left|
      Y_{i,\,\left(t-1\right)}=0, \; X_{i,\,t^\prime\in
      \mathbb{Z}}\right.\right), &\text{if } 0 \le t \le t_{i,\,1}
    \\
    \Prob \left(Y_{i,\,t}= y_{i,\,t} \left| Y_{i,\,\left(t-1\right)}=l, \;
    T_{i,\,1}=t_{i,\,1}, \;\cdots, \;T_{i,\,l}=t_{i,\,l}, \;
    X_{i,\,t^\prime\in \mathbb{Z}}\right.\right), &\text{if } t_{i,\,l} <
    t \le t_{i,\,\left( l+1 \right) }
    \\
    1, &\text{if } t_{i,\,S} < t \ .
    \end{array} \right.
    \label{eq:multfullps}
\end{align}
This result and Equation (\ref{eq:singlefullp}) implies that for each individual
$i$,
\begin{align}
  &\Prob \left( T_{i,\,1}=t_{i,\,1}, \;T_{i,\,2}=t_{i,\,2}, \;\cdots,
  \;T_{i,\,S}=t_{i,\,S} \left| X_{i,\,t^\prime\in \mathbb{Z}} \right. \right)
  \notag \\
    =&\left[\Prob \left( Y_{i,\,t_{i,\,1}}=1\left| Y_{i,\,\left( t_{i,\,1}-1
    \right)}=0,\; X_{i,\,t^\prime\in \mathbb{Z}} \right.\right)
    \prod_{t=0}^{t_{i,\,1}-1}\Prob \left( Y_{i,\,t}=0\left| Y_{i,\,\left( t-1
    \right)}=0, \; X_{i,\,t^\prime\in \mathbb{Z}}\right.\right)\right]
    \cdot
    \notag \\
    & \quad \Bigg\{ \prod_{l=1}^{S-1} \bigg[\Prob \left(
    Y_{i,\,t_{i,\,\left(l+1 \right)}}=l+1\left| Y_{i,\,\left( t_{i,\,\left(l+1
    \right)}-1 \right)}=l,\; T_{i,\,1}=t_{i,\,1}, \;\cdots,
    \;T_{i,\,l}=t_{i,\,l}, \;X_{i,\,t^\prime\in \mathbb{Z}} \right.\right)
    \cdot
    \notag \\
    & \quad
    \prod_{t=t_{i,\,l}+1}^{t_{i,\,\left(l+1 \right)}-1} \Prob \left(
    Y_{i,\,t}=l\left| Y_{i,\,\left( t-1 \right)}=l,\; T_{i,\,1}=t_{i,\,1},
    \;\cdots, \;T_{i,\,l}=t_{i,\,l}, \; X_{i,\,t^\prime\in
    \mathbb{Z}}\right.\right)\bigg] \Bigg\} \ .
    \label{eq:multlikeraw}
\end{align}
With $ l=0, 1,\;\cdots,\;S-1$ we write
\begin{equation}
  \Prob_{i,\,t}\left(l\right)\equiv \left\{ \begin{array}{ll}
    \Prob \left( Y_{i,\,t}=1\left| Y_{i,\,\left( t-1 \right)}=0, \;
    X_{i,\,t^\prime\in \mathbb{Z}}\right.\right), & \text{if } l=0
    \\
    \Prob \left( Y_{i,\,t}=l+1\left| Y_{i,\,\left( t-1 \right)}=l,\;
    T_{i,\,1}=t_{i,\,1}, \;\cdots, \;T_{i,\,l}=t_{i,\,l}, \; X_{i,\,t^\prime\in
    \mathbb{Z}}\right.\right), &\text{if}\:l>0  .
  \end{array} \right.
  \label{eq:multp}
\end{equation}
Since conditional on $Y_{i,\,\left( t-1 \right)}=l$, $Y_{i,\,t}$ can only take
values $l$ or $l+1$, once we get a model for $\Prob_{i,\,t}\left(l\right)$ for
$l=0,\:\cdots,\:S-1$, we can model every term in Equation
(\ref{eq:multlikeraw}). 

Compared with the expression of $\Prob \left( T_i=t_i\left| X_{i,\,t^\prime\in
\mathbb{Z}} \right.\right)$ for a single event (Equation (\ref{eq:likeraw})),
Equation (\ref{eq:multlikeraw}) is much more complicated. In the single event
case, the Markov property implies that, to
model $\Prob \left( T_i=t_i \left| X_{i,\,t^\prime\in \mathbb{Z}} \right.
\right)$  for each individual $i$, it suffices to model the conditional probability
$\Prob\left(Y_{i,\,t}=y_{i,\,t}\left| Y_{i,\,\left(t-1\right)}=0,
\;\mathcal{X}_{i,\,t} \right. \right)$, which is a function of only $t$ and
$\mathcal{X}_{i,\,t}$. However, now we need to model
$\Prob_{i,\,t}\left(l\right)$ for $l=0,\;\cdots,\;S-1$, which is a function of
not only $t$ and $X_{i,\,t^\prime\in \mathbb{Z}}$, but also of $t_{i,\,1},
\;\cdots, \;t_{i,\,l}$, and event state $l$. To simplify this probability
model, We need to make extra assumptions on the dependences among different
events. A simple way is to assume that $\left\{Y_{i,\,t}: t=0,\;1, \;\cdots
\right\}$ is a Markov chain\index{Markov chain}, and then we can proceed just
like the case of single event. However, this assumption may be too restrictive
in many cases. Below, we will provide an alternative approach based on other
assumptions.

\subsection{Regression model}

Assume as above that $Y_{i,\,t}$ only depends on covariate values evaluated at
a finite number of time points, $\mathcal{X}_{i,\,t}$. All the
$X_{i,\,t^\prime\in \mathbb{Z}}$ terms in Equation (\ref{eq:multp}) and on the
RHS of Equation (\ref{eq:multlikeraw}) can then be replaced by
$\mathcal{X}_{i,\,t}$. Also, we write $\{ Y_{i, \, 0}=y_{i, \,0},
\;\cdots,\;Y_{i,\,t}=y_{i,\,t}\}$, $t=0, \;1, \;\cdots$, as $Y_{i,\,0:t}$.

In the Equation (\ref{eq:multp}) for $\Prob_{i,\,t}\left(l\right)$,
$T_{i,\,1}, \;\cdots, \;T_{i,\,l}$, $l=1,\;\cdots,\;S-1$ and covariate vector
$\mathcal{X}_{i,\,t}$ are all conditioning variables. Let us 
treat  $t_{i,\,1}, \;\cdots, \;t_{i,\,l}$ as time-dependent covariates, and
assume an explicit form (with unknown parameters) for
$\Prob_{i,\,t}\left(l\right)$ as a function of $t_{i,\,1}, \;\cdots,
\;t_{i,\,l}$ and $\mathcal{X}_{i,\,t}$.

For example, let $g:\left(0,\,1\right)\rightarrow\left(-\infty, \infty
\right)$ be a monotonic link function\index{link function}, and assume
$g\left( \Prob_{i,\,t}\left(l\right) \right)$ is a linear function or a
polynomial of $t_{i,\,1}, \;\cdots, \;t_{i,\,l}$ and $\mathcal{X}_{i,\,t}$.
For different $l=0,\;\cdots,\;S-1$, the numbers of conditioning event-times
in the expression of $\Prob_{i,\,t}\left(l\right)$ are different. We use a
trick to make the number of covariates constant over time so that our
mathematical expressions can be simply formulated. For each individual $i$, we
define the following time dependent covariates:
\begin{equation}
  T^\prime_{i,\,l} \left(t \right) = \left\{ \begin{array}{cc}
    0, & \text{if } t < t_{i,\,l} \\
    t_{i,\,l}, & \text{if } t \ge t_{i,\,l}
    \end{array} \right., \ \text{for } l=1, \;\cdots,\;S-1 \ .
    \label{eq:multmodel1}
\end{equation}
Now for every $l=0,\;\cdots,\;S-1$, $\Prob_{i,\,t}\left(l\right)$ is a
function of $T^\prime_{i,\,1}, \;\cdots, \;T^\prime_{i,\,\left(S-1 \right)}$,
$\mathcal{X}_{i,\,t}$, $l$ and $t$. If we assume $g\left(
\Prob_{i,\,t}\left(l\right) \right)$ is a linear function of
$T^\prime_{i,\,1}, \;\cdots, \;T^\prime_{i,\,\left(S-1 \right)}$ and
$\mathcal{X}_{i,\,t}$, we can define a covariate vector
\begin{equation}
  Z_{i,\,t} \equiv \left( \mathcal{X}_{i,\,t}^T, \;T^\prime_{i,\,1} \left(t
    \right), \;\cdots, \; T^\prime_{i,\,S-1} \left(t \right) \right)^T \ .
    \label{eq:multmodel2}
\end{equation}
Similarly, if we assume $g\left( \Prob_{i,\,t}\left(l\right) \right)$ to be a
polynomial function of them, we can define $Z_{i,\,t}$ as a vector which
consists of the terms of the polynomial. Under both assumptions, we can write
\begin{equation}
  g\left( \Prob_{i,\,t}\left(l\right) \right) = \beta^T_{t,\,l} Z_{i,\,t},
    \enskip \text{for }l=0,\;\cdots,\;S-1 \ ,
    \label{eq:multmodel3}
\end{equation}
where $\beta_{t,\,l}$ is a parameter vector that varies with time $t$ and
event state $l$ but remains the same across different individuals. In many
situations, we may reasonably assume $\beta_{t,\,l}$ is constant over time.
Then it is a function of only $l$, and we will write it as $\beta_l$.

Now, if all $N$ individuals are independent, and for each individual, we
observe all the $S$ events (i.e. no censoring), then the likelihood function
is
\begin{align}
  L\left(\beta_l\right) &= \prod_{i=1}^N
  \Prob \left( T_{i,\,1}=t_{i,\,1}, \;T_{i,\,2}=t_{i,\,2}, \;\cdots,
  \;T_{i,\,S}=t_{i,\,S} \left| X_{i,\,t^\prime\in \mathbb{Z}} \right. \right)
  \notag \\
  &= \prod_{i=1}^N \Bigg\{ 
  g^{-1} \left(\beta^T_0 Z_{i,\,t} \right)
  \prod_{t=0}^{t_{i,\,1}-1} \left( 1- g^{-1} \left(
  \beta^T_0 Z_{i,\,t} \right) \right)
  \cdot
  \notag \\
  &\qquad \qquad \quad
  \prod_{l=1}^{S-1} \bigg[ g^{-1} \left(
  \beta^T_l Z_{i,\,t} \right) \prod_{t=t_{i,\,l}+1}^{t_{i,\,\left(l+1
  \right)}-1} \left( 1- g^{-1} \left(
  \beta^T_l Z_{i,\,t} \right) \right)
  \bigg] \Bigg\}
  \label{eq:multmodellik}
\end{align}
In the above model, we are making an explicit assumption on the
conditional distribution of $Y_{i,t}$ given all the previous events times. By
successive conditioning, we actually are implicitly making an assumption about
the joint distribution of all the $S$ event-times $T_{i,\,1}, \;\cdots,
\;T_{i,\,S}$ (see Equation (\ref{eq:multlikeraw})). Sometimes, this assumption
may be not easy to verify. On the other hand, even if $\beta_l$ is constant
over $l$, there are $S-1$ more covariates than the single event case. When $S$
is large compared to $N$, the estimates of parameters will have large standard
errors.

\subsection{Estimation and prediction}

We consider the model defined by Equation (\ref{eq:multmodel1}) --
(\ref{eq:multmodel3}). When there is no censoring, the likelihood function is
given by Equation (\ref{eq:multmodellik}). We now turn to the case where the
responses are non-informatively right censored. First, we may assume
$T_{i,\,0}=0$. Note that we have assumed $T_{i,\,l}\neq T_{i,\,l^\prime}$ for
$l\neq l^\prime$ and $l,\,l^\prime=1,\;2,\;\cdots,\;S$ in Section
\ref{sec:notationmultp}. However, it is possible that $T_{i,\,0}=0=T_{i,\,1}$.
For each individual $i$, we observe several times, the last one being the
event-time for the last event or censored time, and the previous times as the
event-times prior to the last observation. If the last observed time is the
time to the last event, there is no censoring; otherwise the observation is
right censored. 

Denote the last observed time by the random variable $\tau_i$ and its value by
$t_i$, where the censoring time is generated by a random variable $C_i$.
Suppose for each individual $i$, prior to the last observed time $t_i$, we
observe $K_i\in \{0,\;1,\;\dots,\;S\}$ events.  Without censoring  $K_i=S-1$,
$\tau_i=t_i=T_{i,\,S}$ and $C_i>T_{i,\,S}$; otherwise, $\tau_i=t_i=C_i$ and
$t_{i,\,K_i} \le C_i < T_{i,\,K_i+1}$. Just as before, we define a censoring
indicator $\delta_i$, which takes values 0 or 1 according as the last
observation is censored or not. Then we can easily show that, under the
non-informative right censoring assumption, the MLE equals the parameter value
that maximizes the following function,
\begin{align}
  L^\prime \left( \beta_l \right) &= \prod_{i=1}^N
    \bigg\{ \Prob \left( T_{i,\,1}=t_{i,\,1}, \;\cdots,
    \;T_{i,\,K_i}=t_{i,\,K_i}, \;T_{i,\,K_i+1}=t_i \left| X_{i,\,t^\prime\in
    \mathbb{Z}} \right. \right)^{\delta_i} \cdot
    \notag \\
    &\qquad \qquad
    \Prob \left( T_{i,\,1}=t_{i,\,1}, \;\cdots, \;T_{i,\,K_i}=t_{i,\,K_i},
    \;T_{i,\,K_i+1}>t_i \left| X_{i,\,t^\prime\in \mathbb{Z}} \right.
    \right)^{1-\delta_i} \bigg\} \ .
\end{align}
The first factor on the RHS of the above equation is 1 when
$\delta_i=0$, and when $\delta_i=1$, it is given by Equation
(\ref{eq:multlikeraw}). When $\delta_i=1$, the second factor is 1 and
when $\delta_i=0$, it is
\begin{align}
  &\Prob \left( T_{i,\,1}=t_{i,\,1}, \;\cdots, \;T_{i,\,K_i}=t_{i,\,K_i},
    \;T_{i,\,K_i+1}>t_i \left| X_{i,\,t^\prime\in \mathbb{Z}} \right.
    \right)
  \notag \\
    =&\prod_{l=0}^{K_i-1} \bigg[\Prob \left(
    Y_{i,\,t_{i,\,\left(l+1 \right)}}=l+1\left| Y_{i,\,\left( t_{i,\,\left(l+1
    \right)}-1 \right)}=l,\; T_{i,\,1}=t_{i,\,1}, \;\cdots,
    \;T_{i,\,l}=t_{i,\,l}, \;X_{i,\,t^\prime\in \mathbb{Z}} \right.\right)
    \cdot
    \notag \\
    &\prod_{t=t_{i,\,l}+1}^{t_{i,\,\left(l+1 \right)}-1} \Prob \left(
    Y_{i,\,t}=l\left| Y_{i,\,\left( t-1 \right)}=l,\; T_{i,\,1}=t_{i,\,1},
    \;\cdots, \;T_{i,\,l}=t_{i,\,l}, \; X_{i,\,t^\prime\in
    \mathbb{Z}}\right.\right)\bigg]
    \cdot
    \notag \\
    &\prod_{t=t_{i,\,K_i}+1}^{t_i} \Prob \left(
    Y_{i,\,t}=K_i\left| Y_{i,\,\left( t-1 \right)}=K_i,\; T_{i,\,1}=t_{i,\,1},
    \;\cdots, \;T_{i,\,l}=t_{i,\,K_i}, \; X_{i,\,t^\prime\in
    \mathbb{Z}}\right.\right)
    \cdot
    \notag \\
    &\qquad\Prob \left( Y_{i,\,0}=0\bigl\lvert X_{i,\,t^\prime\in
    \mathbb{Z}} \right)
    \ .
\end{align}

Once model parameters are estimated, the prediction procedure is not very
different from the case of a single event. We only note here that it will be
computationally challenging to predict all the future events for a new
individual at the same time. Instead, we focus on the time for the next event
conditioned on known previous event-times.

\section{Example} \label{sec:example}

Here, we briefly show an application of our model to a single phenological
event -- blooming of pear trees.

\subsection{Data and objectives}

Representative bloom dates of pear trees in Summerland of British Columbia,
Canada, between 1937 and 1964 were recorded. In each year, a pear tree blooms
at most once, and the bloom date is counted as the number of days from the
first day of a year to a representative bloom date of all the pear trees in
the area under consideration in that year. Note that the time origin ($t_0$)
here is set to January $1^{st}$ of each year. Daily maximum and minimum
temperatures in the same area in the corresponding years are also collected.
It is well known in the agricultural science community that the timing of a
bloom event is closely related to a quantity ``$AGDD$'' -- the accumulation
(cumulative sum) of the so-called growing degree days ($GDD$) defined by
\begin{equation}
  AGDD\left(t\right) = \sum_{k=t_0}^t GDD\left( k \right) \ ,
\end{equation}
where $t_0$ is the time origin, $t$ is the current time (discrete; on daily
scale), and $GDD$ is defined as
\begin{equation}
  GDD\left(k \right) = \left\{ \begin{array}{cc}
    \frac{T_{min}\left(k \right)+T_{max}\left(k \right)}{2} - T_{base} &
    \text{if} \enskip \frac{T_{min}\left(k \right)+T_{max}\left(k \right)}{2}
    > T_{base} \\
    0 & \text{otherwise} \end{array} \right. \;,
    \label{eq:3GDDdef}
\end{equation}
where $k$ is discrete time with the unit of day, $T_{min}\left(k \right)$ and
$T_{max}\left(k \right)$ are daily minimum and maximum temperatures, and
$T_{base}$ is a thresholding constant temperature which is unknown.
Note that (1) $AGDD$ is a function of time; (2)
$T_{base}$ is an unknown parameter; (3) $AGDD$ is not a continuous function of
$T_{base}$. The objective of this data analysis is to predict timings of
future blooming events and to estimate $T_{base}$.

\subsection{Estimation}

Exploratory analysis suggests that the auto-correlation of the bloom dates
over years are negligible. We may therefore assume that these bloom dates on
different years are independent realizations from the same population. We
apply the regression model for single progressive event described in section
\ref{subsec:regModelSingle} to the dataset, using the logit function as link
function. Note that years now play the role of ``individuals''. We assume in
any given year, that on any day, the probability of blooming is only related
to $AGDD$ evaluated at the current time, i.e. that the vector
$\mathcal{X}_{i,\,t}$ contains an intercept and the $AGDD$ value on the
current day. This model then contains three unknown parameters: the intercept,
the coefficient for $AGDD$ evaluated on the current day, and $T_{base}$. The
MLEs of them are: $\hat{\beta}_{intercept}=-22.27$, $\hat{\beta}_{AGDD}=0.07$
and $\hat{\beta}_{T_{base}}=2.97$. A question about these estimators is
whether they are consistent. Wald's \citeyearpar{Wald1949} famous sufficient
conditions for the consistency of MLE requires the likelihood function to be a
smooth function of the parameters. This is not satisfied in our model because
of the presence of $T_{base}$, but we were not able to address this issue
through theoretical analysis.

Instead we explored the issue of consistency through a simulation study. More
precisely, we generated 1000 data pairs of bloom dates and daily average
temperatures of size 30 years, 80 years, 150 years, and 400 years
respectively. We then applied our model to these datasets and calculated the
MLE of each parameter. For each sample size and parameter, we used the average
of 1000 values of the MLE to estimate the mean of the MLE, and their sample
variance to estimate the variance of the MLE. We found that estimated means of
the MLEs get closer to the true parameter values as the sample size increase
from 30 to 400. Moreover, the estimated variances of the MLEs decreases as
the sample size increases. This suggests the MLEs are consistent and gave us
confidence in the value of the estimators.

On the other hand, rather than to rely on the validity of asymptotic theory to
estimate the uncertainties associated with the MLEs, we used bootstrap
confidence intervals. However, the complexity of our model makes it unclear
whether the bootstrap estimates of the quantiles of the MLEs converge to the
true quantiles. We again performed a simulation study to assess that
convergence, the details being similar to those above and hence omitted for
brevity. The results show that the lengths of quantile-based 95\% bootstrap
intervals of the MLEs get very close to those of the estimated 95\% intervals
of the MLEs obtained using the simulated data when the sample size increases.
The bootstrap intervals are slightly biased though (the ends of the bootstrap
interval are always slightly smaller or bigger than the estimated interval
using the simulated data). Overall the results backup use of the bootstrap
intervals to reflect uncertainties in the MLEs. The quantile based 95\%
bootstrap confidence intervals are, for the intercept, (-37.95, -16.62), for
the coefficient for $AGDD\left(t\right)$, (0.055, 0.122), and for
$T_{base}$, (1.93, 3.81). We see that the both the intercept and the
coefficient for $AGDD\left(t\right)$ differ significantly from 0 at the 5\%
level.

\subsection{Prediction}

To use (\ref{eq:preddistplugin}) or (\ref{eq:predictivedMC}) to predict the
representative bloom date of the pear trees of the next year in Summerland, we
need to predict the daily average temperature ($\left(T_{min}\left(t
\right)+T_{max}\left(t \right)\right)/2$) of that year first. After removing
seasonality in daily historical temperature data, We fit an $ARIMA$ model to
the residual historical temperature. Note the deseasonalized historical
temperature series contain weak periodic signals with longer periods, and
therefore is not strictly stationary. Nevertheless, $ARIMA$ may still be used
as a a reasonable approximation. By comparing Bayesian information criterion
(BIC) for $ARIMA$ models of different orders, we settled on $ARIMA\left(3, \,
0, \, 1\right)$ as our final model for generating future temperatures in any
given growing season.

We now turn to prediction. At the end of the current year, we generate 1000
series of the daily average temperatures of the whole next year using the
fitted $ARIMA\left(3, \, 0, \, 1\right)$ model. We then use
(\ref{eq:predictivedMC}) to calculate the probability of the blooming event
happening on each successive day of the following year. This way we get a
(discrete) predictive distribution for the timing of that blooming event.

So suppose that we are on the end of the first day of the new year with its
observed average daily temperature. We then apply the $ARIMA$ model to
generate 1000 temperature series starting from the second day of the new year.
As above, we can use (\ref{eq:predictivedMC}) to get another predictive
distribution for the timing of the blooming event. We repeat this procedure on
each successive day, until the true bloom date, at which time prediction
ceases. If the true bloom date is around day 129, we then get 129 successive
predictive distributions. What we expect to see are increasingly more accurate
predictions as the days progress toward the bloom date and more and more
information about the daily averages temperatures come to hand for that
season. Growing confidence in that prediction would provide an increasingly
strong basis for management decisions.

To see if our expectations are realized, we performed a leave-one-out
prediction procedure -- at every step, leave out one year of data for
assessment and use the remaining years for training the model to predict the
bloom date in the left-out year.  For each left-out year, we follow the
prediction scenario described above. As a result we get 28 years (1937--1964)
of assessments, with a total of 3523 predictive distributions, the average of
the bloom dates for those years being about day 126. For each of these
predictions, we calculate the median of the predictive distribution as a point
prediction of the new bloom date. Along with that we calculate a
quantile-based 95\% prediction interval (PI) for the new bloom date. With all
the 3523 predictive distributions, we then can estimate the root mean square
error (RMSE) and mean absolute error (MAE) of the prediction, as well as the
coverage probability of the 95\% PI. The results are as follows: the RMSE is
5.65 days; the MAE is 4.36 days; the estimated coverage probability of the
95\% PI is 99\%; the average length of the PIs is 30 days. 

The coverage probability of the 95\% PI is too high, plausibly because in the
$ARIMA(3,\,0,\,1)$ model we have incorporated in the random noise term, the
variability in the temperature series caused by deterministic periodic signals
other than seasonal variation. In any case, reducing the variance of the white
noise in the $ARIMA(3,\,0,\,1)$ model by half of its estimated value yields
improvement and we get: the RMSE is 5.79 days, the MAE is 4.33 days, the
estimated coverage probability is about 94\%, and the average length of the
PIs is 21 days.

\begin{figure}[!ht]
\begin{center}
    \includegraphics[scale=.6]{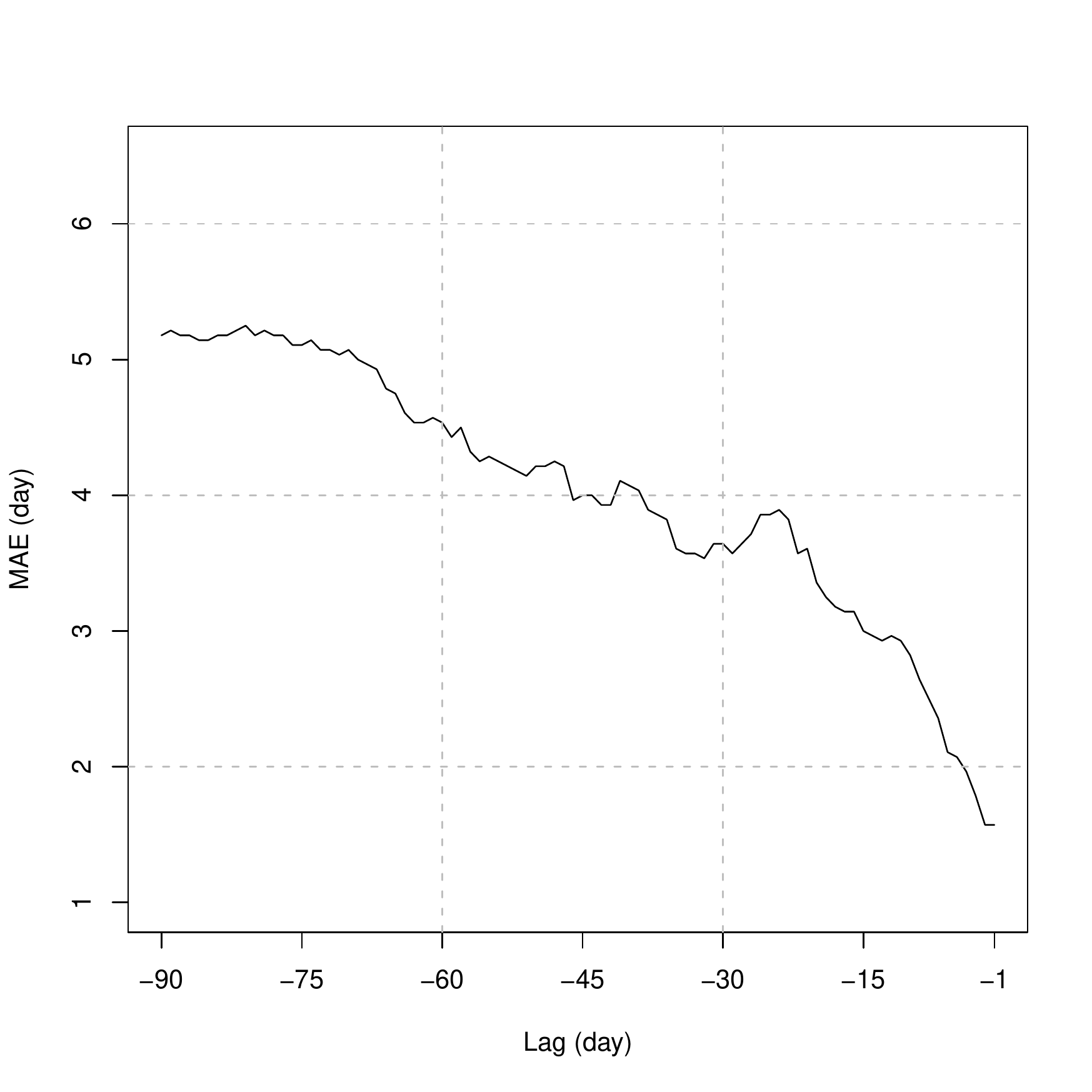}
    \caption{Change of the MAE with the change of lag. The point
    prediction becomes more accurate when time approaches the bloom
    date.}
    \label{fig:uncMAE}
\end{center}
\end{figure}

\begin{figure}[!ht]
\begin{center}
    \includegraphics[scale=.6]{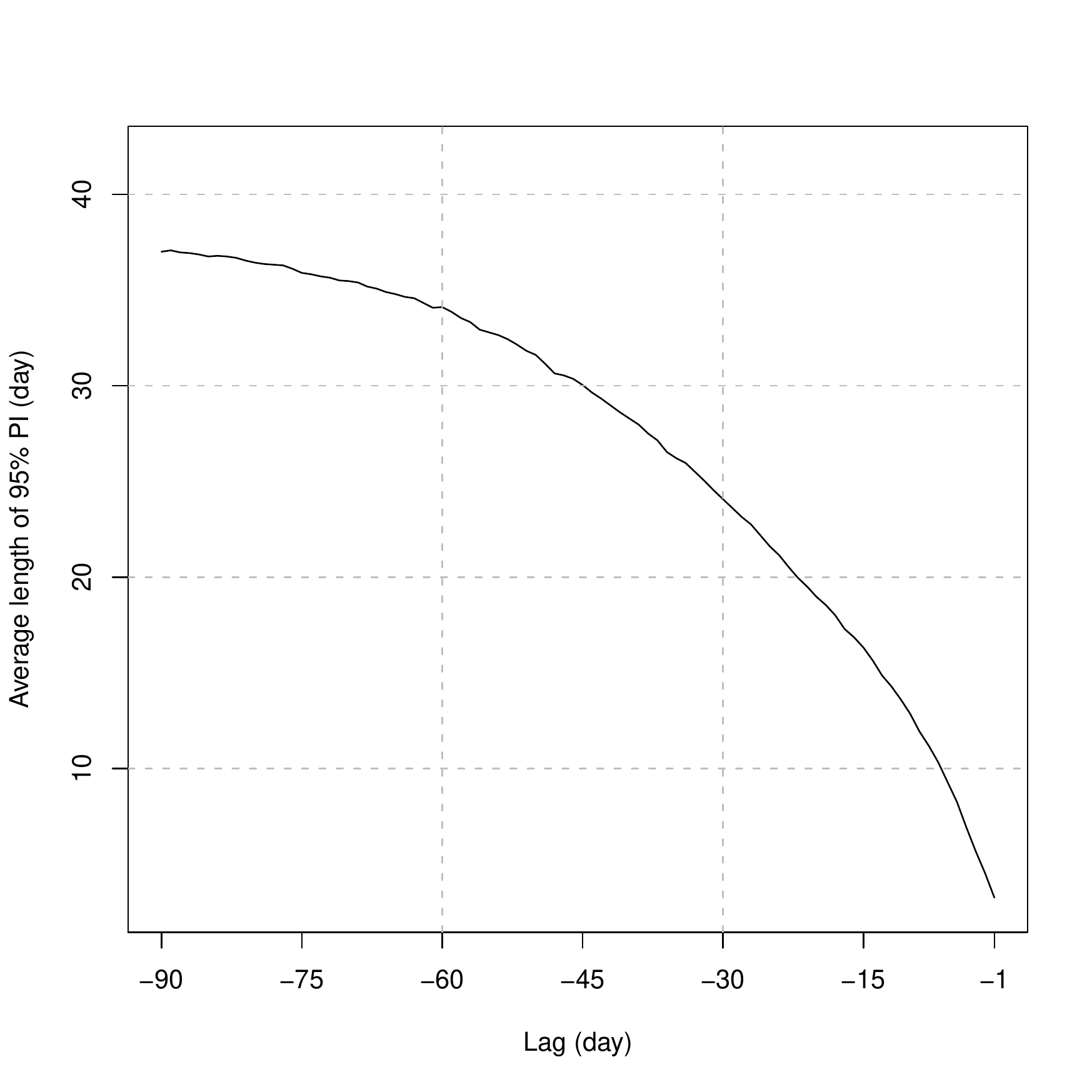}
    \caption{Change of the average length of 95\% PIs with the change of lag.
    The predictive uncertainty decreases when time approaches the bloom
    date.}
    \label{fig:uncPI}
\end{center}
\end{figure}

As noted above, we expected the prediction to become more accurate as time
approaches the real bloom date. To check this, we calculated the MAE and the
average length of the 95\% PIs each day over the years of interest, beginning
90 days prior to the bloom date (call it ``lag -90'') to 1 day prior to the
bloom date (``lag -1''). The results for the MAE and the average length of the
95\% PIs are shown in Figure \ref{fig:uncMAE} and \ref{fig:uncPI}
respectively. We see that the MAE does become smaller and the average length
of the 95\% PIs, shorter as the actual bloom date approaches in line with our
expectations. In fact, by the time we reach one month prior to the bloom date,
the prediction has become quite accurate (the MAE is about 3.5 days).

The above results for prediction are influenced by two models: one is our
regression model for a single progressive event, the other is a crude ARIMA
model for daily average temperature. To check the pure performance of our
regression model, we performed the leave-one-out procedure again. But this
time, we assume all the future daily average temperatures are known. Note
that, in this case, we cannot give a sensible estimate for the coverage
probability of the 95\% PIs since for each test year, we can only get one
predictive distribution. The results are very good: the RMSE is 2.64 days, the
MAE is 1.89 days, and the average length of the 95\% PIs is 9.21 days.
Although this is no longer a real prediction, these results tend to validate
our regression model for the blooming event. This finding also demonstrates
the importance of modeling the covariate series accurately and points to the
need of improving the temperature forecasting models.

\section{Concluding Remarks} \label{sec:conclusin}

The regression models presented in this paper aim at the prediction of the
times of progressive events when time-dependent covariates that are known up
to discrete time points are present. Instead of directly modeling the hazard
function, we model the process of the binary state indicators. This way, all
the time-dependent information can be easily incorporated by considering a
model for a binary variable at each time point. When there is only a single
event, the process of the state indicators is a Markov chain. But when there
are multiple events, that process does not necessarily have a Markovian
structure. In this case, some additional assumptions are needed for
simplifying the probability model and circumventing computation challenges
that would otherwise arise. Application of our approach to bloom date data has
shown that the prediction using it can be quite accurate. Although originally
designed for phenological data, these models should be useful for a broad
range of survival data.

A restrictive distributional assumption in our models is that the process of
the state indicator needs to be time-homogeneous. One way to relax this
assumption might be to allow the model parameters to change with time. Another
restrictive assumption is that in the multiple events case, we require that no
two events can happen at the same time point. However, in practice, this may
occur, especially when the discrete time scale is coarse. We will need some
further work to remove this restriction.

\subsection*{Acknowledgements}

We acknowledge Dr. Denise Neilsen (AAFC-Summerland Research Station) for
providing phenology data for testing of our regression model. We thank the
Canadian Government Research Affiliate Program (RAP) and ”Growing Forward”
federal government funding of the Sustainable Agriculture Environmental
Systems (SAGES) research program to Agriculture and Agri-Food Canada (AAFC)
for stipend support of Mr. Song Cai and research support for (Dr, N.
Newlands). Partial financial support for the research was produced by grants
from Canadian National Institute for Complex Data Structures (NICDS) and the
Natural Science and Engineering Research Council (NSERC). Song Cai was also
supported by the Alexander Graham Bell Canada Graduate Scholarship (CGSM) from
NSERC from September 2010 -- April 2010.

\bibliographystyle{plainnat}
\bibliography{bibmodelingEvent.bib}

\end{document}